\documentclass[18pt]{article}
\usepackage{epsfig}
\def\apj{ApJ}
\def\apjl{ApJL}
\def\mnras{MNRAS}

\def\etal{{\it et al.}}
\def\d{{\bf d}}

\def\kn{k_{\parallel}}

\def\u{{\bf U}}

\def\th{{\vec{ \theta}}}

\def\d{{\bf d}}
\begin{document}
\title{HI Fluctuations at Large Redshifts: III - Simulating the  Signal
  Expected  at GMRT. }
\author{Somnath Bharadwaj \footnote{Corresponding author, \, \,
  email:\,  somnath@phy.iitkgp.ernet.in }
\\ Department of Physics and Meteorology \\ \& \\ 
  Center for Theoretical Studies,\\
I.I.T. Kharagpur, 721 302, India \\ \\
Pennathur Sridharan Srikant   \footnote{Present address: Department of
  Physics, University of Utah, USA} 
\\ Department of Physics and Meteorology \\ 
I.I.T. Kharagpur, 721 302, India }
\date{}
\maketitle
\baselineskip=18pt
\begin{center}
Abstract
\end{center}
We simulate the distribution of  neutral hydrogen (HI) at the 
redshifts $z=1.3$ and $3.4$ using a cosmological N-body simulation
along with  a prescription for  assigning HI masses to the
particles. The HI is distributed in clouds whose properties are
consistent with those of the damped Lyman-$\alpha$ absorption systems
(DLAs) seen in quasar spectra. The clustering properties of these
clouds are identical to those of the dark matter. We use this to
simulate the redshifted HI emission expected at $610 \, {\rm MHz}$
and $325 \, {\rm MHz}$, two of the observing bands a the  GMRT. These  
are used to  predict the correlations expected between the complex
visibilities measured at different  baselines and frequencies in
radio-interferometric  observations with the GMRT.  The visibility
correlations directly probe the power spectrum of HI fluctuations at
the epoch when the HI emission originated, and this holds the
possibility  of 
using HI observations to study large-scale structures at high $z$. 
\newpage
\section{Introduction}
Observations  of Lyman-$\alpha$ absorption lines  seen in quasar
spectra are an important probe of  the distribution of
neutral hydrogen (HI) at high redshifts. 
These observations show that the bulk of the neutral gas in the
redshift range $1 \le z \le 3.5$ is in HI clouds with column    
densities greater than $2 \times 10^{20} {\rm atoms/cm^{2}}$ (Peroux
{\it et al.} 2001, Storrie-Lombardi, McMahon, Irwin 1996, Lanzetta, Wolfe,
\& Turnshek 1995). The damped Lyman-$\alpha$ absorption lines produced
by these clouds  indicate $\Omega_{gas}(z)$, the comoving  
density of neutral gas expressed as a fraction of the present critical 
density, to be  nearly constant at a value   $\Omega_{gas}(z) \sim
10^{-3}$  (Peroux {\it et al.} 2001). 

In this paper we simulate the  HI emission expected from these
 clouds. The aim of the exercise is to 
investigating the possibility of detecting the
 redshifted HI   emission  using the Giant Meterwave Radio Telescope
 (GMRT; Swarup et al., 1991). We focus 
 on two of  the GMRT frequency bands centered at $610 \,{\rm MHz}$ and
 $325   \,{\rm MHz}$ corresponding  to HI emission from redshifts
 $z=1.3$ and  $3.4$.    The HI flux from individual clouds 
($ < 10 \mu {\rm Jy}$) is too weak to be detected by GMRT 
unless the image of the cloud is   significantly
magnified  by an intervening cluster gravitational lens (Saini,
 Bharadwaj and Sethi, 2001). Although we may not be able to detect 
individual clouds, the redshifted HI emission from the distribution of  
clouds will appear as  a background radiation in  low frequency radio 
observations. In  three earlier papers ( Bharadwaj, Nath \& Sethi,
 2001; Bharadwaj  \& Sethi, 2002;  and Bharadwaj \& Pandey,  2003; 
 hereafter referred to as Paper a, b and c respectively), and in
 the  present paper we investigate issues related to calculating  the  
 expected signal and detecting it. 

We  propose (Papers b and c ) that the optimal observational 
strategy for detecting this signal is  to deal directly with the
complex visibilities measured in radio interferometric
observations. Briefly introducing the terms involved, we remind the
reader that the quantity measured in radio interferometric
observations with an array of antennas is the complex visibility
$V(\u,\nu)$. This is measured for every pair of antennas at every
frequency channel in the observation band.  For every pair of
antennas it is convenient to express the visibility as a function of
$\u=\d/\lambda$ {\it ie.} the separation between the two antennas $\d$ 
expressed in units of the wavelength $\lambda$. We refer to the
different possible values of $\u$ as baselines. 
 One of the big advantages of dealing directly with the visibilities
 is that the system noise contribution to   the visibilities is
 uncorrelated. The visibilities respond  only to   the fluctuations in
 the redshifted HI emission. In Paper b we showed  
that the  correlation expected between the visibilities  $V(\u,\nu)$
and  $V(\u,\nu+\Delta \nu)$ measured at the same baseline at two
slightly different frequencies is 
\begin{eqnarray}
\langle V(\u,\nu) V^{*}(\u,\nu+\Delta \nu) \rangle =
\frac{[\bar{I}b D  \theta_{0}]^2}{2 r^2}
\int_0^{\infty} d \kn  P(k) \left[1+ \beta \frac{\kn^2}{k^2}
  \right]^2 \cos(\kn  r^{'} \Delta \nu) 
\label{eq:a0}
\end{eqnarray}
where  $ k=\sqrt{(2  \,   \pi \, U/r)^2  +   \kn^2}$, 
$\bar{I}_0$ is the specific intensity expected from the HI
emission if the HI were uniformly distributed, 
$\theta_0=0.6 \times \theta_{\rm FWHM}$,  $\theta_{\rm FWHM}$ being
the angular width of the primary beam of the individual antennas, $r$
is the comoving distance to the HI from which the radiation
originated, and  $b^2 D^2 P(k) \left[1+ \beta   \frac{\kn^2}{k^2}
  \right]^2$  is the 
power spectrum of the fluctuations in the HI distribution in redshift
space at the epoch  when the HI emission originated.  

To summarize, the  visibility-visibility cross-correlation (hereafter
refereed to as the visibility correlation) directly  probes 
the  power spectrum of HI  fluctuations at the epoch where the HI 
emission originated. This holds the possibility of allowing us to study
the large scale structures at high redshifts. A point to note is that
the the visibility correlations at a baseline $U$ receives
contribution from the power spectrum only for Fourier mode $k>k_{\rm
  min}=(2 \pi/r) U$, and for the CDM-like power spectrum most of the
contribution comes from Fourier modes around $k_{\rm min}$. So, it
may be said that the correlations at a baseline $U$ probes the power
spectrum at the Fourier mode $(2 \pi/r) U$. 

In the earlier work we treated the HI as being continuously
distributed whereas in reality the HI resides in discrete  gas
clouds. In addition, it was assumed that the HI distribution is an
unbiased representation of the underlying dark matter distribution,  and
we used the linear theory of density perturbations to follow the
evolution of fluctuations in the dark matter distribution. It is these
assumptions which allow us to express the power spectrum of
fluctuations in the HI distribution in redshift space  at the epoch
when the HI emission  originated (in eq. \ref{eq:a0}) in terms of 
$b$ the linear bias 
parameter (taken to be 1), $D$ the growing mode of linear density 
perturbations (Peebles, 1980) at the epoch when the HI emission
originated, $P(k)$ the present power spectrum of dark matter density
fluctuations calculated using linear theory and the factor  $\left[1+
  \beta   \frac{\kn^2}{k^2}   \right]^2$ which takes the power
spectrum from real space to 
redshift space in the linear theory of redshift distortions  (Kaiser 1987) .   
 Here we report progress on two counts. First, we have used a  PM
 N-body code to evolve the fluctuations in the dark matter
 distribution, thereby incorporating possible non-linear
 effects. Second, we have assigned HI 
masses to the dark matter particles  in the N-body code and this 
was used to simulate the redshifted HI emission. So we have also  been
able to incorporate the fact that the HI gas is contained in discrete
clouds. The predictions for the HI signal expected at GMRT  presented
in this paper incorporate both these effects.  We still retain the
assumption that the HI is an unbiased tracer of the dark matter.

We next present a brief outline of this paper.  In section 2 we
discuss the method that was used to simulate the HI signal, and in
Section 3. we present the results of our investigations. In Section 4
we discuss the results and present conclusions.  

Finally, it should be point out that there been alternative lines of
approach investigating the possibility of using HI observations to
study large scale structures at $z \sim 3$ (Sunyaev \& Zeldovich,
1975; Subramanian \&  Padmanabhan, 1993; Kumar, Padmanabhan \&
Subramanian, 1995; Weinberg et al., 1996; Bagla, Nath \&  Padmanabhan,
1997; Bagla \& White, 2002). The reader is referred to Papers a and b
for a detailed comparison of these approaches with that adopted here.  
\section{Methodology.}
We have simulated the visibility correlations expected at
two of the GMRT observing frequency bands centered at $\nu_c=610 \,
{\rm   MHz}$ and $325 \, {\rm MHz}$.  The simulations were carried out
in three steps 
\begin{itemize}
\item[1.] Using a PM N-body code to simulate  the dark matter
  distribution  at the redshift where the HI emission originated
\item[2.] Assigning HI masses to  the  particles used
  in the N-body code and calculating the flux expected from each HI cloud  
\item[3.] Calculating the complex visibilities arising  from the
  distribution of HI clouds and
  computing the  visibility correlations. 
\end{itemize}

We next discuss the salient features of each of these steps.
The values  $h=0.7$, $\Omega_{m0}=0.3$ and $\Omega_{\Lambda0}=0.7$
were used throughout. 

\subsection{The N-body Simulations.}
We have used a Particle-Mesh (PM) N-body code to simulate the dark
matter distribution at the redshift $z$ where the HI emission
originated. 
The simulation volume was a cubic box of comoving volume $L^3$. The
size $L$ was chosen so that it is approximately twice  the comoving
distance subtended by $\theta_{{\rm FWHM}}$ of the GMRT primary beam. 

The values of $r$  the comoving distance to the region from where the
HI emission originated, the grid spacing of the mesh $\Delta L$, and
the  number of dark matter particles used in each simulation $N_{\rm
DM}$ are all shown in Table~1.   

\begin{center}
\begin{table}{Table 1.}\\  \\
\baselineskip=18pt
\begin{tabular}{|c|c|c|c|c|c|c|c|c|}
\hline 
$\nu_c$ (MHz) &  $z$ & $\theta_{{\rm FWHM}}$ & $r$ (Mpc) & $L$ (Mpc) &
$\Delta L$ (Mpc) & $N_{{\rm DM}}$ & $N_{\rm SIM}$ & $z_{in}$\\
\hline
610 & 1.33 & $0.9^{\circ}$ & 4030 & 128 & 0.5 & $128^3$ & 4 & 19\\
325 & 3.37 & $1.8^{\circ}$ & 6686 & 512 & 1 & $256^3$ & 4 & 9 \\
\hline
\end{tabular}
\end{table}
\end{center}

The initial power spectrum of dark matter density fluctuations at
$z_{in}$ (shown in Table~1) is normalized to COBE (Bunn \& White
1996), and its shape 
is  determined using the   analytic fitting form for the CDM power
spectrum given by Efstathiou, Bond and White (1992). The value of the
shape parameter turns out to be $\Gamma=0.2$ for the set of
cosmological parameters used here. We have run the N-body code for
$N_{\rm SIM}$ (Table 1) independent realisations of the initial
conditions and the final results for the visibility
correlations were  averaged over all the realisations. 

\begin{figure}[htb]
\begin{center}
\mbox{\epsfig{file=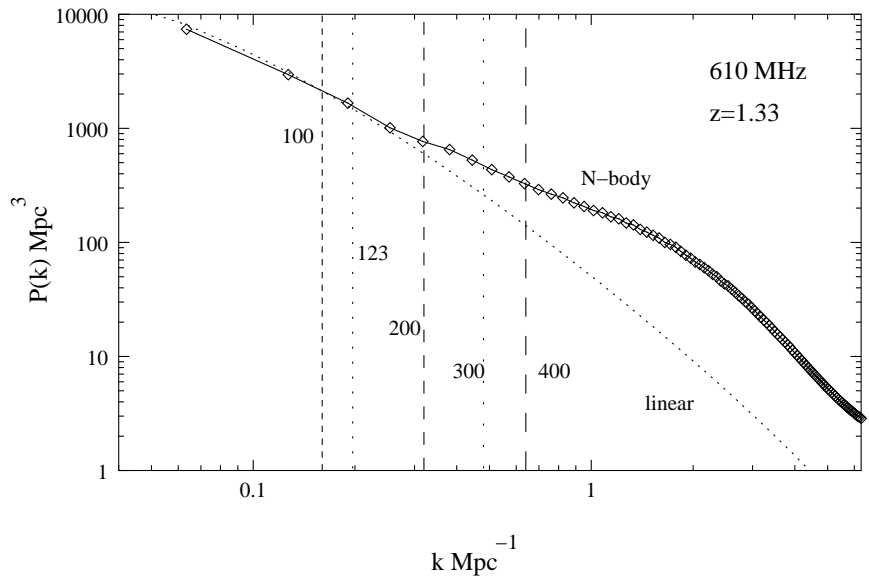,width=4.5in,angle=0}}
\caption{\label{fig:2}\baselineskip=18pt
This shows the power spectrum  of density fluctuations in the dark
matter distribution at $z=1.33$. The vertical lines show the smallest
Fourier mode $k_{\rm  min}=(2 \pi/r) U$ which contributes to the
visibility correlations  at a baseline $U$. This is shown for  the
different values of $U$ indicated  in the figure.}
\end{center}
\end{figure}

\begin{figure}[hbt]
\begin{center}
\mbox{\epsfig{file=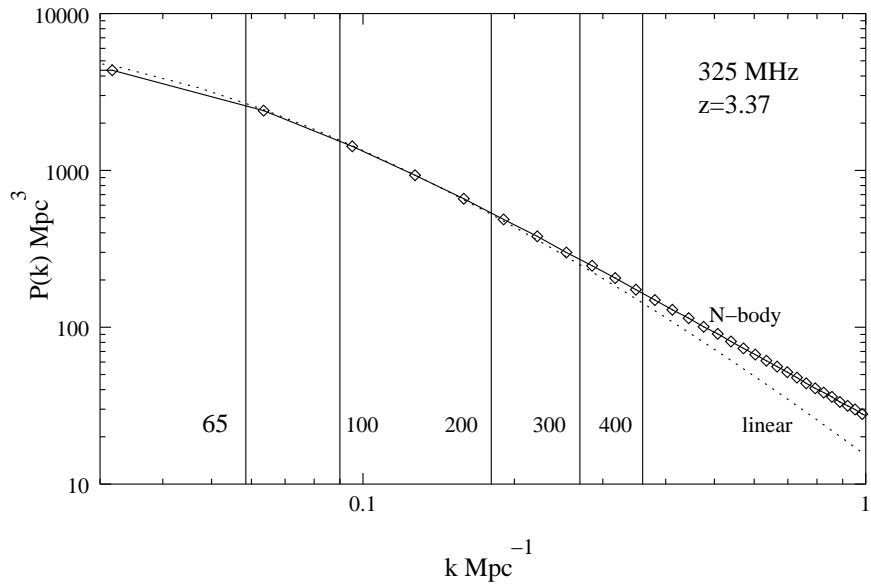,width=4.5in,angle=0}}
\caption{\label{fig:3} \baselineskip=18pt
This is the same as figure 2 except that it is at $z=3.37$ which
corresponds to $325 {\rm MHz}$.} 
\end{center}
\end{figure}

The N-body code gives the final positions and peculiar velocities of
the $N_{\rm DM}$ dark matter particles in the simulation. The 
power spectrum of the density fluctuations in the dark matter
distribution at $z=1.33$ and $z=3.37$ are  shown in Figures
\ref{fig:2} and \ref{fig:3} 
respectively. In both the figures  we have
shown the power spectrum  for the  range of Fourier modes which will
make a significant contribution  to the visibility correlation at the
baselines where the signal  is expected 
to be strongest. We find that at $z= 1.33 \, (610 {\rm MHz})$ the
power spectrum obtained from the N-body simulation shows substantial
differences from the power spectrum calculated using linear theory at
Fourier modes $k \ge 0.3 \, {\rm Mpc}^{-1}$. Converting to baselines,
we expect non-linear effects to be important for $U \ge 200$. At
$z=3.37 \, (325 {\rm MHz})$ there 
are differences between the N-body and linear power spectrum at $k\ge
\, 0.3 {\rm Mpc}^{-1}$, but the non-linear effects are not as
pronounced as 
at $325 {\rm MHz}$. Converting to baselines, non-linear effects will
influence the signal at baselines $U \ge 400$. 

\subsection{Assigning HI masses.}
We model the HI clouds as randomly oriented rotating disks of radius
$R$, column density $N_{\rm HI}$ and  rotation velocity $V$. The
values of $R$ and $V$ are held fixed in each simulation, and we have
run simulations with different sets of values for these parameters. 
It is assumed that the column densities   have a power law 
distribution  in the range  $2 \times 10^{20} \le N_{\rm HI}/({\rm
atoms/cm^{2}}) \le 1 \times 10^{22}$ and the comoving number
density of HI clouds with column densities in the interval $d N_{\rm
  HI}$ is $B N_{\rm HI}^{-\alpha} \, d N_{\rm HI}$.  The total comoving
number density of HI clouds at redshift $z$  is 
\begin{equation}
n_{\rm HI}^c(z)=B \, \int^{N_{\rm HI}[{\rm max}]}_{N_{\rm HI}[{\rm
      min}]}  N_{\rm HI}^{-\alpha} \, d N_{\rm HI}
\label{eq:a1}
\end{equation}
and the comoving mass density is 
\begin{equation}
\rho_{\rm HI}^c(z)=B \, \int^{N_{\rm HI}[{\rm max}]}_{N_{\rm HI}[{\rm
      min}]} \, (\pi R^2 N_{\rm HI} m_{\rm HI}) 
N_{\rm HI}^{-\alpha} \, d N_{\rm HI}
\label{eq:a2}
\end{equation}
where $m_{\rm HI}$ is the mass of the hydrogen atom. The  normalisation
coefficient $B$ is determined by using  equation (\ref{eq:a2}) to
calculate $\Omega_{gas}(z)$
\begin{equation}
\Omega_{gas}(z)=\frac{4}{3} \Omega_{\rm HI}(z) =\frac{4}{3}
\frac{8 \pi G }{ 3 H^2_0} \rho^c_{\rm HI}(z)
\end{equation}
We use $\Omega_{gas}=10^{-3}$ in all the simulations.

\begin{center}
\begin{table}{Table 2.}\\  \\
\begin{tabular}{|c|c|c|c|c|}
\hline
$\nu_c$ (MHz) &  $\alpha$ & $R$ Kpc  & $n_{\rm HI}^c(z) \, {\rm
  Mpc}^{-3}$  & $N_{\rm clouds}$\\
\hline
610 & 1.2 & 10 & $2.0 \times 10^{-2} $ & 42467 \\
610 & 1.2 & 8 & $3.2 \times 10^{-2} $ & 66354 \\
610 & 1.2 & 5 & $8.1 \times 10^{-2} $ & 169865 \\
610 & 1.2 & 2 & $5.1 \times 10^{-1} $ & 1061651  \\
610 & 1.7 & 10 & $3.7 \times 10^{-2} $ & 76764 \\
610 & 1.7 & 8 & $5.7 \times 10^{-2} $ &  119943 \\
610 & 1.7 & 5 & $1.5 \times 10^{-1} $ & 307055   \\
610 & 1.7 & 2 & $9.2 \times 10^{-1} $ &  1919088 \\
325 & 0.8 & 10 & $1.3 \times 10^{-2} $ & 1799234 \\
325 & 1.2 & 10 & $2.0 \times 10^{-2} $ & 2717825 \\
325 & 1.2 & 8 & $3.2 \times 10^{-2} $ & 4246602  \\
325 & 1.2 & 5 & $8.1 \times 10^{-2} $ & 10871300 \\
\hline
\end{tabular}
\end{table}
\end{center}

This model fixes the total number of HI clouds in the simulation
volume $N_{\rm clouds}=L^3 n_{\rm HI}^c(z)$ . The total
number of clouds scales as 
\begin{equation}
N_{\rm clouds} \propto \frac{1-\alpha}{2-\alpha} R^{-2}
\end{equation}
as  we vary  $\alpha$ the slope of the column density distribution
 or the radius of the clouds $R$. For large values of $R$ the HI is 
distributed in a few clouds with large masses, whereas  there are many
clouds with low HI masses when $R$ is small.  Our model has three free
 parameters,  namely $\alpha$, $R$ and $V$. We have run simulations
 varying  $\alpha$ and $R$ (Table 2) for $V=100 {\rm  \, km/s} \, {\rm and}
\,  200 \, {\rm  km/s}$.

We randomly select $N_{\rm clouds}$ particles from the output of the
N-body simulation and these are identified as HI clouds.  The HI mass
of each cloud  is $M_{\rm   HI}=\pi R^2 N_{\rm HI} m_{\rm HI}$, where
the column density is drawn randomly from the power-law
distribution discussed earlier. The center of the simulation volume is
aligned with the center of the GMRT primary beam and it is 
located at a comoving distance corresponding to the  redshift $z$. The
comoving distance to each cloud  is used to calculate its angular
position and redshift. The redshift   is used to determine the 
luminosity distance which is used to calculate the  flux from the
individual clouds. The effect of the  peculiar velocity is
incorporated when calculating $\nu_o$ the frequency at which the HI
emission from each cloud is received. The line width $\Delta \nu$ of
the HI emission line from each cloud is calculated using $\Delta
\nu= \mid \sin \theta \mid \, 2\nu_o  V/c $, where $2\nu_o  V/c $ is
the line width if the disk of the galaxy  were viewed edge on and
$\theta$ is the angle between the normal to the disk and the line of
sight.    

To summarize, at the end of this stage of the simulation we have
$N_{\rm clouds}$ HI clouds. For each
cloud  we have  its angular position $\vec{\theta}^{a}$,and the  flux  
density $F^a$, frequency $\nu^a_o$ and line-width $\Delta \nu^a$
of the redshifted HI emission. Here the index $a$ ($1 \le a \le
N_{\rm clouds}$) refers to the different clouds in the simulation. 

\subsection{Calculating visibility correlations}
We first describe how we have calculated the complex visibilities that
would be measured in GMRT radio observations of the  HI
distribution generated in the simulation. 
The observations are  carried out at  $NC$ frequency channels $\{
\nu_1,\nu_2,\nu_3, ...,\nu_{NC}\}$ covering a frequency band   $B$
centered at the frequency $\nu_c$. We have used $B=8 {\rm MHz}$ and
$NC=64$ at  $\nu_c=610 {\rm MHz}$, and $B=8 {\rm MHz}$ and
$NC=128$ at  $\nu_c=325 {\rm MHz}$.  

 For the purpose of this paper we   
assume that the  antennas are distributed on a plane, and that they all
point  vertically up wards. The beam pattern $A(\th)$  quantifies how the
individual antenna, pointing up wards,  responds to signals from
different directions in the sky. This is assumed to be a Gaussian
$A(\th)=e^{-\theta^2/\theta_0^2}$ where $\theta_0=0.6 \times 
\theta_{\rm FWHM}$ (Table 1).  

The position of each antenna can be denoted  by a two dimensional vector 
$\d_i$.    The quantity  measured in interferometric  observations is 
the visibility $V(\u,\nu)$ which is recorded for every independent
pair of antennas (baseline) at every frequency channel in the
band. For any pair of antennas, the visibility depends on the vector
$\d =\d_i - \d_j$  joining the position of  the two antennas. It is
convenient  to  express the visibility as a function of the variable
$\u$ which is $\d$    expressed  in units of the  wavelength {\it
  i.e.} $\u=\d/\lambda$.  The signal arising from the
clustering pattern of the HI clouds will be strongest at the small
baselines, and  our calculations  have  been limited to this. We have
considered a square grid of baselines extending from $-U_{\rm max}$ to 
$U_{\rm  max}$ with resolution  $\delta U$. We have used $U_{\rm
  max}=400$ and $\delta U=10$.  The  complex visibility
has been calculated for  each baseline $\u$ on the grid  using 
\begin{equation}
V(\u,\nu)=\sum_{a=1}^{N_{\rm clouds} } \, A(\vec{\theta}^a) \, F^a
e^{- i 2 \pi \u \cdot \th^a} O(\frac{\mid \nu-\nu_o^a \mid}{\Delta
  \nu^a})
\end{equation}
where the function $O(x)$ is defined such that $O(x)=1$ for $x\le1$,
else $O(x)=0$.
It is to be noted that in an actual GMRT observation the baselines  
will have a complicated distribution depending on which part of the
sky is  observed and the duration of the observation. Given the
fact that the signal we are interested in is statistical in nature,
and   that we are interested in making generic predictions about the
signal expected in a typical GMRT observation, a square grid of
baselines is adequate. 

The final step in the simulation is to calculate the
visibility correlation $\langle V(\u, \nu) V^{*}(\u,
\nu+\Delta \nu) \rangle$. The angular  brackets $\langle 
\rangle$ indicate the ensemble average, and we have averaged over the 
$N_{\rm SIM}$ different realisation of the N-body simulation. In
addition, the correlation depends only on the separation in frequency
$\mid \Delta \nu \mid$, and the magnitude $U=\mid \u \mid$.  So, for a 
fixed values of $\Delta \nu$ and $U$ we have averaged  over all
possible pairs of frequencies and baselines which match these values. 

The analytic calculations (Papers b and c) where the HI is assumed
to have a continuous  distribution, predict the imaginary part of the
visibility correlation function to be zero, and the
clustering signal is manifest in only the real part. In the
simulations we get a very small, but non-zero imaginary 
component. This is not discussed in the rest of the paper where we
present results for the real component only.   

\section{Results}
In this section  we present results for the visibility
correlation  as  obtained from our simulations. We compare these with
the analytic predictions of Papers b and c and investigate the
effect of two factors (1.) the non-linear evolution of the density
fluctuations, and (2.) the discrete nature of the HI
distribution. To get a better understanding of the second 
effect, we present results varying the parameters of the HI
distribution. 

\subsection{610 {\rm MHz}}

\begin{figure}[h]
\begin{center}
\mbox{\epsfig{file=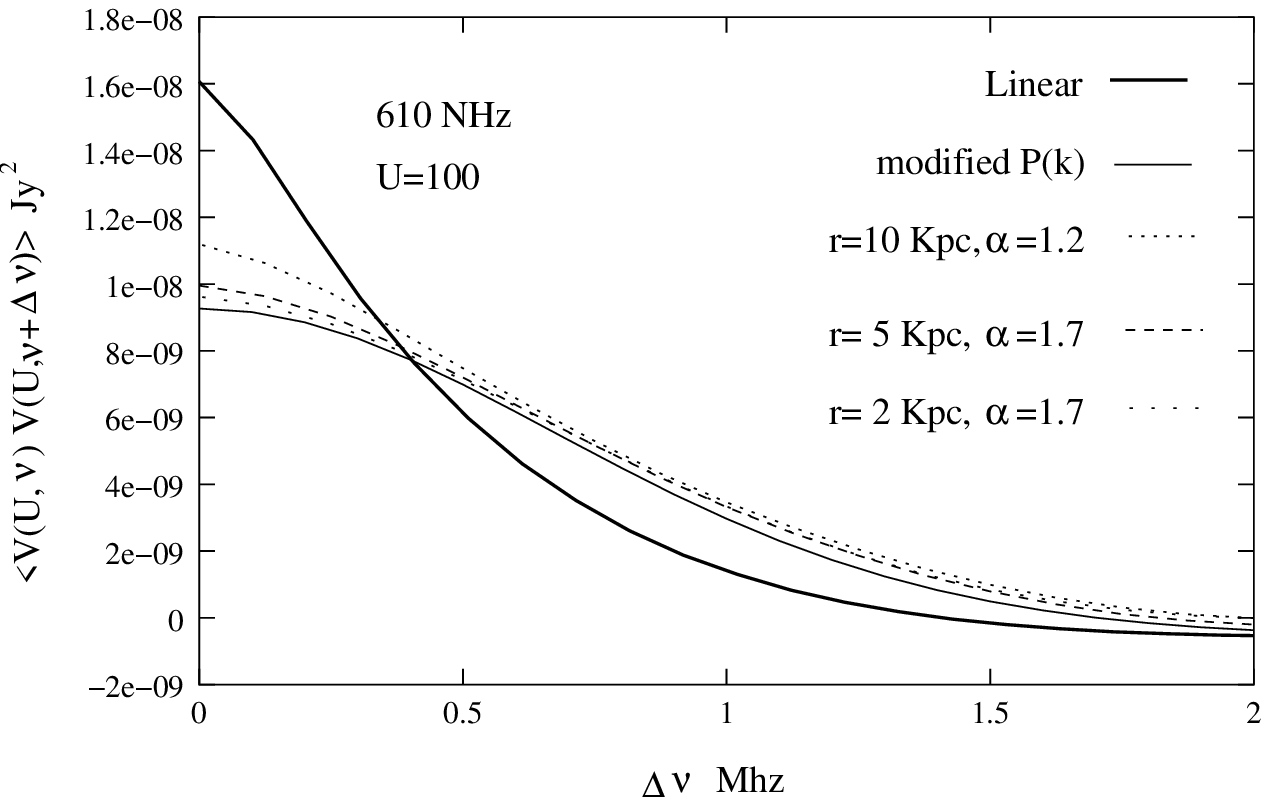,width=4.5in,angle=0}}
\caption{\label{fig:4} \baselineskip=18pt
This shows the correlation expected  between the visibilities
$V(\u,\nu)$ and $V(\u,\nu+\Delta \nu)$  at the same 
  baselines at two different  frequencies. The rotational velocity of
  the HI disk is assumed to be $V=200 \, {\rm km/s}$. The other 
parameters of the HI   distribution take on values shown in the
figure.  These results are for the $610   {\rm MHz}$ band.}
\end{center}
\end{figure}

\begin{figure}[hbt]
\begin{center}
\mbox{\epsfig{file=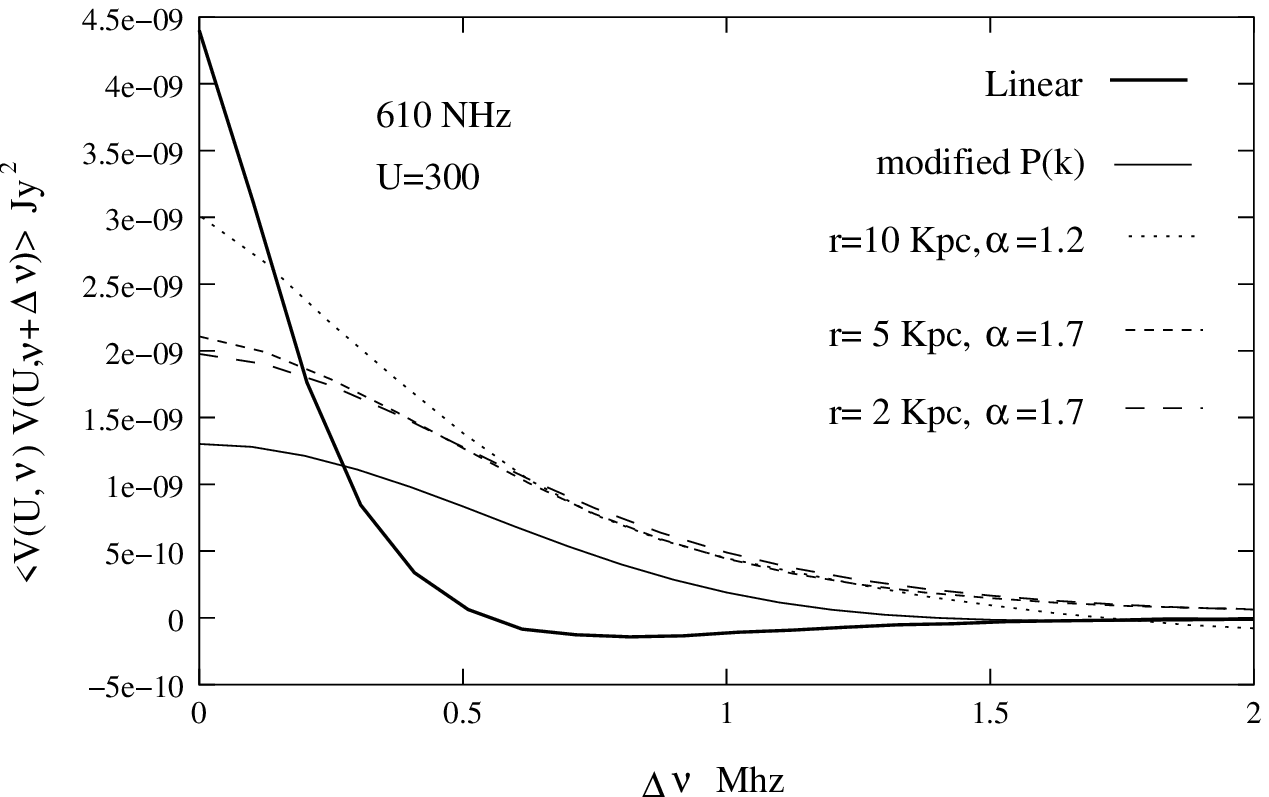,width=4.5in,angle=0}}
\caption{\label{fig:7}
\baselineskip=18pt
 This is the same as Figure \ref{fig:4} except
  that it shows results for  $U=300$.  }
\end{center}
\end{figure}

Figure \ref{fig:4} shows the visibility correlations 
for $U=100$, the results at smaller baselines show a similar
behaviour. The visibility correlation at the baseline $U=100$ receives  
contributions mainly  from Fourier modes around $k\sim 0.2 \, {\rm
  Mpc^{-1}}$  which is in the linear regime (Figure \ref{fig:2}),  and
we expect 
a good agreement with the analytic, linear predictions. We find that
for $\Delta \nu < 0.5{\rm MHz}$ the analytic 
predictions are larger than the correlations obtained in the
simulations, and this is reversed for  $\Delta \nu > 0.5 {\rm
  MHz}$.  This discrepancy can be explained if we take into account the
fact that visibility correlation actually responds to the clustering in
redshift space. It is known  (e.g. Suto \& Suginohara 1991; Graman,
Cen \& Bahcall, 1993;  Fisher et  al. 1994; Brainerd et al., 1996; 
Bromley, Warren \& Zurek, 1997)  that non-linear effects
can be important in redshift space even on scales where the  
clustering in real space is well described by linear perturbation
theory. It has been shown that this can be modeled by taking into
account the effect of the random motions along the line of sight (e.g. Fisher
\etal 1994, Peacock \& Dodds 1994, Ballinger, Peacock \& Heavens 1996).   
We incorporate this by multiplying the power spectrum with
$\exp[-\kn^2 \sigma^2]$ in our analytic formulas for the visibility
correlation. This gives the modified formula
\begin{eqnarray}
\langle V(\u,\nu) V^{*}(\u,\nu+\Delta \nu) \rangle &=&
\frac{[\bar{I} b D \theta_{0}]^2}{2 r^2}
\int_0^{\infty} d \kn  P(k) \left[1+ \beta \frac{\kn^2}{k^2} 
  \right]^2 \times \nonumber \\ &&\exp[-\kn^2 \sigma^2]
\cos(\kn  r^{'} \Delta \nu) 
\label{eq:a8}
\end{eqnarray}
for the visibility correlation. 
We find that for $\sigma=200 \,{\rm km/s}/H_0$ this gives a good fit
to the results of the simulations,  and this is also shown in the
figure.    
We next shift our attention to how the results depend on the
parameters of the HI distribution. We find that for $r=10 \, {\rm Kpc}$
where the bulk of the HI is distributed in a few clouds with large HI
masses the results show a $20 \%$ increment at  small values of
$\Delta \nu$  compared to the  models with smaller values of $r$. This
excess correlation at small 
$\Delta \nu$ arises from the fact that the HI emission from an
individual cloud will be spread across a width $\delta \nu$ in
frequency.  The correlation between the HI emission from the same HI
cloud at two different frequencies  will contribute to the visibility 
correlations when $\Delta \nu \le \delta \nu$. The 
contribution from this signal is significant  in comparison to that
arising 
from the clustering of the HI clouds when the total HI is distributed
in a few clouds with large HI masses each. The contribution to the
visibility correlation from within individual HI clouds goes down as
$r$ is reduced and the HI is distributed among many clouds each with
small HI masses. There is very little difference between the results
for $r=5 {\rm Kpc}$ and $2 {\rm Kpc}$ and we may treat this as the
result if the HI were continuously distributed.     

The results in Figure \ref{fig:4}  are for the rotational velocity
  $V=200 \,{\rm   km/s}$.  We have also done simulations using 
$V=100 \,{\rm   km/s}$. We
  find that there are   differences ($<20 \%$) only at  
  small   values of $\Delta \nu$. The effect of decreasing $V$ is to
  decrease   the frequency width of  the HI emission line from
  individual   clouds which results in a higher value of the HI flux
  density. This   does not effect the clustering signal but enhances   
the contribution to the visibility   correlation arising from the
  emission of   a single HI cloud.  As changing $V$ does not affect
  the results very much, in this subsection we show the
  results for $V=200 \,{\rm   km/s}$ only.

Figure  \ref{fig:7}  shows the results for
$U=300$. We find that the discrepancy between the linear, 
analytic predictions and the results of our simulations increases  at
larger values of  $U$. Except at very small values of $\Delta \nu$, the
simulated values are larger than the linear predictions. This is
because the larger baselines probe smaller length scales which are
significantly nonlinear (figure \ref{fig:2}) and the amplitude of the
fluctuations is larger than predicted by linear theory. 
At smaller scales the fluctuations are non-linear even in
real space,  and the modified formula (equation \ref{eq:a8}) based on
only redshift space considerations grossly underestimates the
visibility correlations.  An important point is that at larger values of
$U$ the visibility  correlations calculated in the simulations do not
fall as sharply with increasing $\Delta \nu$ as predicted in the
linear calculations. In Paper c we found that the visibility
correlations decay as  $\propto 
\exp[-\Delta \nu/K]$,  where the decay constant varies as $K\propto  
U^{-0.8}$ {\it i.e.} the decay is faster at larger baselines. Our
simulations show that the decay with increasing $\Delta \nu$ is 
slower 
than predicted using linear theory. This is a consequence
of the fact that the density fluctuations are non-linear on the
length-scales being probed at these baselines. Another point to note
is that the dependence on the parameters $R$ and $\alpha$, becomes
relatively more pronounced at large values of $U$.  

\subsection{325 {\rm MHz}}

\begin{figure}[htb]
\begin{center}
\mbox{\epsfig{file=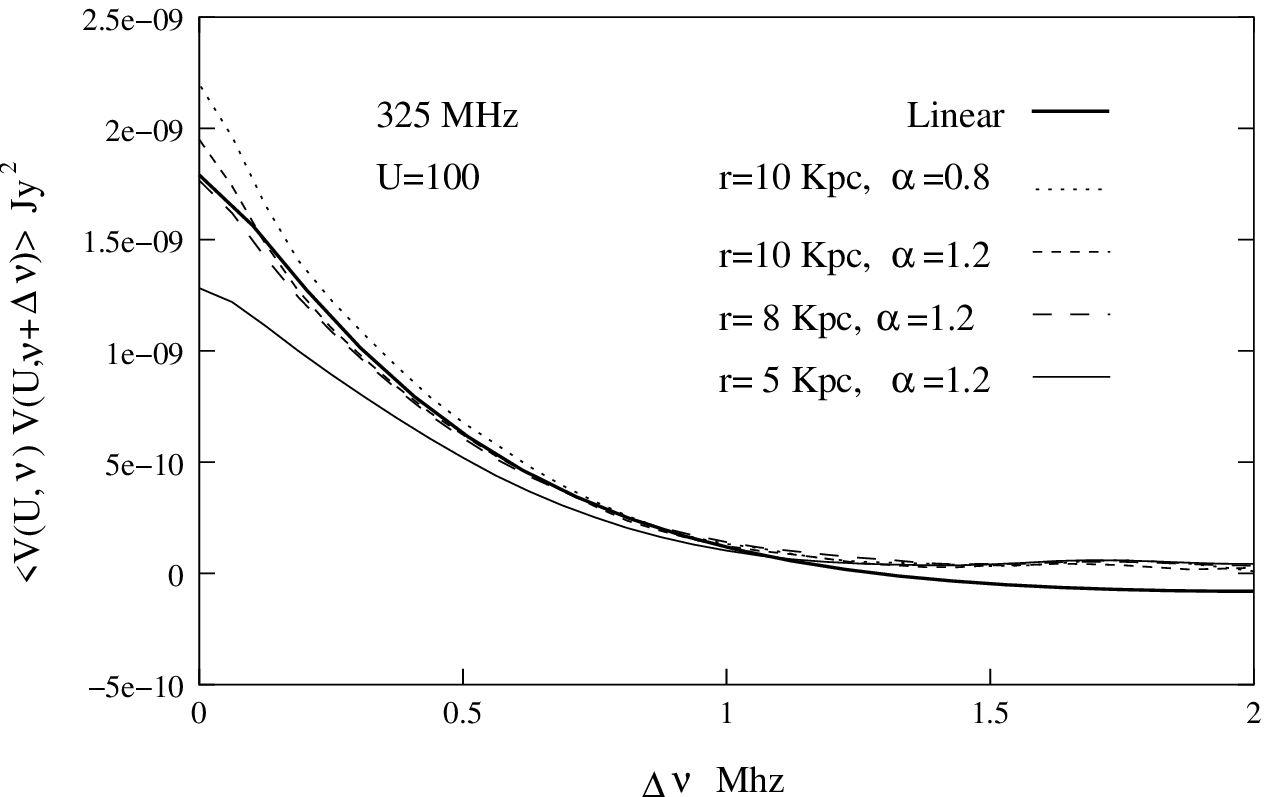,width=4.5in,angle=0}}
\caption{\label{fig:9}
\baselineskip=18pt
This shows the expected correlation between the visibilities
$V(\u,\nu)$ and $V(\u,\nu+\Delta \nu)$  at the same 
baselines at two different  frequencies. The rotational velocity of
the HI disk is assumed to be $V=100 \, {\rm km/s}$. The other 
parameters of the HI   distribution take on values shown in the
figure.  These results are for the $325   {\rm MHz}$ band.}  
\end{center}
\end{figure}

\begin{figure}[hbt]
\begin{center}
\mbox{\epsfig{file=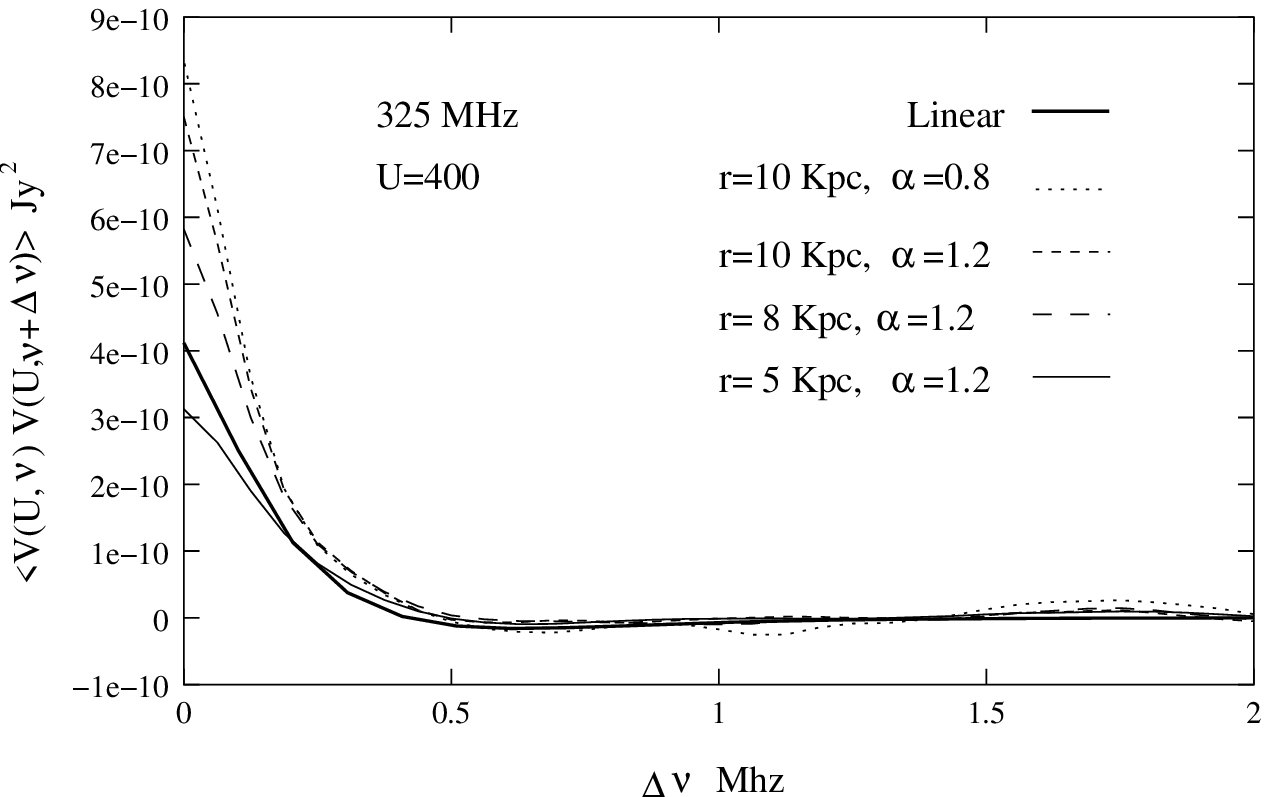,width=4.5in,angle=0}}
\caption{\label{fig:12}
\baselineskip=18pt
This is the same as Figure \ref{fig:9} except that it shows
  results for $U=400$.
}
\end{center}
\end{figure}

A point which should be mentioned right at the start is that at $325
{\rm MHz}$ we are restricted in the values of $R$  for
which we are able to carry out simulations. At $325 {\rm MHz}$ the
simulation volume is pretty large (Table 1) and for $R=2 {\rm Kpc}$
the total number of HI clouds in the simulation volume becomes too
large  for our  computational resources.  Also, in this subsection we
use $V=100 \, {\rm km/s}$ in our simulations. 
Figure \ref{fig:9}  shows the results for the visibility correlations
at $U=100$. The behaviour at smaller baseline`es is not very
different. We find that for $\Delta \nu <1 {\rm MHz}$ the predictions
of the  analytic, linear  calculations (Paper c) are  
very close to the values obtained in  the simulation for $r=8 {\rm 
  Kpc}$ and $\alpha=1.2$. The results of the simulation are slightly
larger than the analytic predictions when $r=10 {\rm Kpc}$,  and they
are somewhat smaller than the analytic predictions at $r=5 {\rm Kpc}$.
For $\Delta \nu >1 {\rm   MHz}$ the results of the simulation are the
same for all the parameters,  and  the value is slightly more than the
analytic prediction.  The power spectrum (Figure {\ref{fig:3}) is in
  the linear 
regime at the Fourier modes which contribute to the
visibility correlations at $U=100$. The discrepancy between the
analytic predictions and the results of our simulations can be
attributed to a combination of the  two factors discussed earlier (1.)
the effect of random motions on the redshift space clustering and (2.)
correlations between the HI emission from the same cloud  at different
frequency channels.  

Figure \ref{fig:12} shows the
visibility correlations at $U=400$.  We find that the  behaviour of
the visibility  correlations does not change very much 
for  baselines in the range $100 < U \le 400$.    The
power spectrum  (Figure \ref{fig:3}) starts getting
non-linear at Fourier modes corresponding to $U=400$, but the effect
is not very significant. 

\section{Discussion and Conclusions.}
We take up for discussion  two issues pertaining to
the way we have modeled the distribution of HI clouds.  First
is our  assumption that 
the HI clouds responsible for damped Lyman-$\alpha$ absorption
lines are rotating disks, all with the same radius and rotational
velocity.   Prochaska and  Wolfe (1998) have proposed that these  HI
clouds are the gaseous progenitors of present day  galaxies. They have     
attempted to explain the observed kinematic of the damped
Lyman-$\alpha$ absorption lines using a thick, rotating disk  model
for the HI clouds. Another model  ( Haehnelt, Steinmetz  \&Rauch,
1998) proposes that the observations could be better explained 
by modeling the absorption systems as protogalactic gas clumps
undergoing merger. The second issue is our assumption that the HI
column 
densities have a power law distribution and our choice of the value of
the index  $\alpha$. Lanzetta {\it et al.} (1991) show that the 
column density distribution at $z \simeq 2.5$ can be described by a
power law with $\alpha \simeq 1.7$. In a later paper Lanzetta, Wolfe
and Turnshek (1995) show that the column density distribution evolves
quite strongly with redshift, there being a tendency toward more high
column density clouds at higher redshifts. In our work we have run
simulations for two values of the index {\it i.e.}  $\alpha=1.2$ and
$1.7$.  Having very briefly reviewed some of the prevalent views and
having compared our assumptions with them, we note that our
simulations seem to indicate that the visibility correlation signal
does not depend very critically on the details of the properties of
the HI clouds. We find that the visibility correlation signal 
has contribution from mainly two effects (1.) correlations caused by
the emission from the same cloud to different frequency channels,
and (2.) the clustering of the clouds in redshift space. The first
effect is seen in the correlation at small values of $\Delta \nu$, 
where $\Delta \nu$ is smaller than the width of the HI line from an
individual cloud. This effect manifests itself as a  rise in the
visibility correlations at small values of $\Delta \nu 
\, (< 0.5 {\rm MHz})$. This effect is enhanced  if the HI is distributed  
in a few clouds with large HI masses as compared to the situation
where the HI is in many small clouds with low HI masses. This 
seems to be the only effect of the fact that the HI is distributed in
discrete HI clouds and is not continuously  distributed. This is also
the only place where the details of the HI distribution affects the
visibility correlation.    

We next turn our attention to the contribution to the visibility
correlation signal from the clustering of the HI clouds. We may take
the results for the visibility correlations at the values of the
parameters $R$ and $\alpha$ where $N_{\rm clouds}$ is maximum as
representing the results when the discrete nature of the HI
distribution can be neglected (continuum limit). This assumption is  
justified at $610 \,{\rm MHz}$ where the is very little difference  in
the  results between $R=5\, {\rm Kpc}$ and $R=2\, {\rm Kpc}$.  
We have not carried out simulations for $R<5\, {\rm
  Kpc}$  at $325 \, {\rm MHz}$,  and simulations with smaller values of
$R$ are needed at this frequency before we can be sure that the 
results for $R=5 \, {\rm Kpc}$ really represent the continuum limit.  
Let us first discuss the results at the baselines  for which the
visibility correlation probes the power spectrum  at length-scales
which are  in the linear regime. This is true  for baselines with $U
\le 100$ at $610 \, {\rm MHz}$ (Figure \ref{fig:2}). At $325\, {\rm
  MHz}$ most of the  baselines which we have studied probes
the power spectrum in the linear regime (Figure \ref{fig:3}). For all
these baselines we find that  the
simulated values are less than the predictions of linear theory at
small $\Delta \nu$ and the simulated values are larger than the linear
predictions at large $\Delta \nu$. The transition occurs in the range
$\Delta \nu \sim 0.5-1 \, {\rm MHz}$. 
We propose  that this discrepancy
is a consequence of the fact that the fluctuations in the HI
distribution in redshift space may be non-linear even on length-scales
where linear theory holds in real space. This can be modeled by
incorporating the effect of random peculiar velocities on the redshift
space HI distribution. We show that including this effect gives a good
fit to the simulated results at $U=100$ for $610 \, {\rm MHz}$. At
larger baselines the visibility correlation probes the power spectrum
on length-scales where it is non-linear. Non-linear effects start
influencing the visibility correlation at  baselines $U\ge 200$ 
for $610 \, {\rm MHz}$, and these effects are very significant  by
$U=400$. As a consequence of these effects the simulated visibility
correlations do not fall of with increasing $\Delta \nu$ as quickly as
predicted by linear theory. Also, the simulated values are larger than
the linear predictions everywhere except at very small values of
$\Delta \nu$. The range of $\Delta \nu$ where the simulated value are
less than the linear predictions decreases with increasing $U$.    

In conclusion we note that the HI signal predicted by our simulations
 are not drastically different from the analytic
predictions presented earlier. In this paper we have been able to
address the effects of the discrete nature of the HI distribution and
the non-linear nature of the HI fluctuations  in redshift space. We
now have the tools necessary to simulate the HI signal expected at the
GMRT. A  full simulation of a GMRT observation requires us to also include
the system noise as well as  various galactic and extragalactic radio 
sources. Only then will we be able to make definite predictions 
as to whether it will be possible to detect the HI signal or not. Work 
is correctly underway on this. The preliminary results indicate that
 it will be posiible to have a $5 \, \sigma$ detection at $610 \,
 {\rm MHz}$ with  one thousand  hours of observation.

{\it Acknowledgment.} SB would like to than Jasjeet S Bagla,  Jayaram
N Chengalur and  Shiv K Sethi for useful discussions. SB would also
like to acknowledge BRNS, DAE, Govt. of India,for financial support
through sanction No. 2002/37/25/BRNS.  
\newpage
References
\begin{itemize}
\item[] Bagla J.S., Nath B. and Padmanabhan T. 1997, MNRAS 289, 671
\item[] Bagla J.S. and White M. 2002, astro-ph/0212228
\item[] Ballinger W.E., Peacock J. A., Heavens A. F., 1996,
\mnras, 282, 877 
\item[] Bharadwaj S., Nath B. \& Sethi S.K. 2001, JAA. 22, 21
\item[] Bharadwaj, S.~\&  Sethi, S.~K.\ 2001, JAA, 22, 293  
\item[] Bharadwaj, S, \&  Pandey, S. K., 2003, JAA, 24, 23
\item[] Brainerd T. G., Bromley B. C., Warren M. S.,
Zurek W. H., 1996, \apj, 464, L103 
\item[] Bromley B. C., Warren M. S., Zurek W. H., 1997,
\apj, 475, 414  
\item[] Bunn E. F. \&  White M. 1996,  ApJ, 460, 1071
\item[] Efstathiou, G., Bond, J. R.  \& White, S. D. M. 1992,
 MNRAS, 250, 1p
\item[] Fisher K. B., Davis M., Strauss M. A., Yahil A., Huchra
J. P., 1994, \mnras, 267, 927
\item[] Gramman M., Cen R., Bahcall N. A., 1993, \apj, 419, 440
\item[] Haehnelt, M.~G., Steinmetz, M., \& Rauch, M., 1998, ApJ, 495, 647 
\item[]  Kaiser N., 1987, MNRAS, 227, 1
\item[]  Kumar A., Padmanabhan T. and Subramanian K., 1995, MNRAS, 272,
      544
\item[] Lahav O., Lilje P. B., Primack J. R. and Rees M., 1991, MNRAS, 251, 128
\item[] Lanzetta, K. M., Wolfe, A. M., Turnshek, D. A. \& Lu, L.,  1991,
    ApJS, 77, 1
\item[] Lanzetta, K. M., Wolfe, A. M., Turnshek, D. A. 1995,
    ApJ, 430, 435
\item[] Peacock J. A., Dodds, S.J., 1994, \mnras, 267, 1020
\item[] Peebles, P. J. E. 1980, {\it The Large-Scale Structure of
    the Universe \/}, Princeton, Princeton University Press
\item[] P\'eroux, C., McMahon, R. G., Storrie-Lombardi,
    L. J. \& Irwin, M .J. 2001, MNRAS 
\item[] Prochask, J. X. \& Wolfe, A. M., 1998, ApJ, 507, 113
\item[]  Saini, T., Bharadwaj, S. \& Sethi, K. S. 2001, ApJ, 557, 421
\item[] Storrie--Lombardi, L.J., McMahon, R.G., Irwin, M.J. 1996, MNRAS,
   283, L79
\item[] Subramanian K. and Padmanabhan T., 1993, MNRAS, 265, 101
\item[] Sunyaev R.A. and Zeldovich Ya.B., 1975, MNRAS, 171, 375
\item[]Suto Y.,  Suginohara T., 1991,\apjl, 370, L15    
\item[] Swarup, G., Ananthakrishan, S., Kapahi, V. K., Rao, A. P.,
 Subrahmanya, C. R., \& Kulkarni, V. K. 1991, Curr. Sci., 60, 95
\item[] Weinberg D.H., Hernquist L., Katz N.S. and Miralda-Escude J.,
      1996, Cold Gas at High Redshift eds. M.Bremer, H.Rottgering,
      C.Carilli and P.van de Werf, Kluwer, Dordrecht

\end{itemize}

\end{document}